\documentstyle[prb,multicol,aps]{revtex}
\begin{document}
\def\dla{\langle\!\langle}
\def\dra{\rangle\!\rangle}

\draft

\title{ Spin chirality induced by the Dzyaloshinskii-Moriya interaction
 \\ and the polarized neutron scattering}
\author{D.N. Aristov, S.V. Maleyev}
\address{ Petersburg Nuclear Physics Institute,
Gatchina, St. Petersburg 188350, Russia}\date{\today} \maketitle

\begin{abstract}
We discuss the influence of the Dzyaloshinskii-Moriya (DM) interaction
in the Heizenberg spin chain model for the observables in the polarized
neutron scattering experiments.
We show
that different choices of the parameters of DM interaction may leave
the spectrum of the problem unchanged, while the observable spin-spin
correlation functions may differ qualitatively. Particularly, for the
uniform DM interaction one has the incommensurate fluctuations and
polarization-dependent neutron scattering in the paramagnetic phase.
We sketch the possible generalization of our treatment to higher
dimensions.
\end{abstract}



\begin{multicols}{2} \narrowtext


Since the works by Dzyaloshinskii and Moriya \cite{DzyaMo}, the
antisymmetric spin exchange interaction plays an important role in the
physics of condensed matter. Being introduced for the explanation
of the weak ferromagnetism in antiferromagnets without center of
inversion, the Dzyaloshinskii-Moriya (DM) interaction is
found nowadays in various problems of magnetism and
statistical physics.

Being the relativistic effect, the magnitude of DM interaction, $D$, is
generally expected to be small in comparison with the usual symmetric
superexchange, $J$.\cite{DzyaMo,SEWA} In some compounds, however, this
interaction can attain a sizeable value. For instance, one has $D/J =
0.18$ in the hexagonal perovskite CsCuCl$_3$ \cite{SKS} and $D/J
\approx 0.05$ in copper benzoate. \cite{Aff-Osh}

Remarkably, the DM interaction in the latter compounds takes place
in the quasi-one-dimensional spin subsystems. On this reason,
we are primarily concerned below with the one-dimensional
(1D) situation of a quantum spin chain.

Generally, the DM interaction between two spins, ${\bf S}_{1,2}$, is
written as ${\bf D} ({\bf S}_1 \times {\bf S}_{2} )$ with
an axial DM vector ${\bf D}$. In a chain, ${\bf D}$ may spatially
vary both in direction and magnitude, however, the symmetry
arguments usually rule out most of the possibilities and confine the
theoretical discussion to two principal cases. The first one is the
uniform DM interaction, ${\bf D}=const$ over the system. \cite{SKS}
The second case is the staggered DM interaction,\cite{Aff-Osh} with
antiparallel ${\bf D}$ on adjacent bonds.

Among the other studies, we should mention the discussion of
the $XY$ spin chains with randomly distributed values of $D$ \cite{De-Ri}
and the growth models with imaginary uniform $D=i\lambda$ leading to
non-Hermitian Hamiltonian. \cite{growth} A model of $XY$
spin chain with a ternary DM interaction was introduced and
solved recently.  \cite{ternary}

In a present paper, we deal mostly with two above cases, uniform and
staggered DM interaction. We consider also a model of a non-ideal
lattice, where one finds, say, an almost uniform situation with one
possible DM
value, $\bf D$, taking place on a chain fragment of average length
$l_1$, and another value, $-\bf D$, on a chain fragment of length
$l_2$, while $l_{1,2}\gg1$. The situation is then described in
probabilistic terms.  We show that, being the situation uniform,
staggered or random, the spectrum of the Heisenberg chain with the DM
interaction is equivalent to one of XXZ spin model and is computed
exactly for spin 1/2. Of course the observable susceptibilities can
differ crucially, as we demonstrate below.

Therefore we extend the previous result by Alcaraz and
Wreszinski, that the Heisenberg 1D model with the uniform DM
interaction is reduced to XXZ spin exchange model and is exactly
solvable.  \cite{Alc-Wre}

For the uniform and almost uniform DM interaction, we show that
the observable spin-spin correlation function possess an incommensurate
structure. This incommensurability phenomenon was noted
previously for the $XY$ spin chain model and uniform
antisymmetric spin interaction.
\cite{Ko-Tsu}
We show further that in this case the spin susceptibility tensor
$\chi^{\alpha\beta}$ acquires an antisymmetric part.
This leads to the appearance of the polarization-dependent part of
the neutron scattering cross-section, which makes possible the
direct observation of the direction and the value of the DM vector
${\bf D}$. Our results are applicable in the absence
of the long-range magnetic order in the system.  They also can be
generalized towards higher-dimensional situation, as discussed below.
Therefore our treatment may provide an explanation of the earlier
experiments in cubic ferromagnet MnSi, \cite{MnSi} where the DM-induced
incommensurability of the magnetic fluctuations and polarization
dependence of the neutron scattering were observed both below and above
the Curie ordering temperature.


We consider the spin chain Hamiltonian of the form

        \begin{eqnarray}
        {\cal H }  &=&
        \sum_{l=1}^{L}( J
        {\bf S}_l {\bf S}_{l+1}
        + {\bf D}_l [{\bf S}_l \times {\bf S}_{l+1} ])
        \label{iniHam}
        \end{eqnarray}

\noindent
with AF Heisenberg coupling $J$ and the Dzyaloshinskii-Moriya term
${\bf D}_l$.
We choose the vector ${\bf D}_l$ to be directed along the $z-$axis.

We observe that ${\cal H }$ is simplified upon
a canonical
transformation ${\cal H} \to e^{-iU} {\cal H} e^{iU}$ with

        \begin{equation}
        U = \sum_{l=1}^L \alpha_l S_l^z,  \qquad
        \alpha_l  = \sum_{i=1}^{l-1} \tan^{-1} (D_{i}/J),
        \label{canon}
        \end{equation}
and $\alpha_1 =0$.
Note that this transformation works for all values of $S$.
The periodic boundary conditions (BC) require $\alpha_{L+1} = 0\, {\rm
mod}\, 2\pi$ which relation is not generally satisfied. However, in the
thermodynamic limit $L\to\infty$ the influence of BC can be
negliged. \cite{Alc-Wre}
Introducing $S_j^\pm = S_j^x \pm i S_j^y$,
one can easily see that

        \begin{eqnarray}
        {\widetilde S}_l^{\pm} &\equiv&
        e^{-iU} S_l^\pm e^{iU} = S_l^{\pm}
        e^{\mp i\alpha_l }, \quad
        {\widetilde S}_l^z
        =S_l^z ,
        \label{newspins}
        \end{eqnarray}

\noindent
and our choice of the coefficients $\alpha_l$ removes the
antisymmetric part of the Hamiltonian,
$[{\widetilde {\bf S}}_l \times {\widetilde {\bf S}}_{l+1}] $.
We consider below two
principal possibilities: the ``uniform'' situation $ D_l = J
\tan\delta$ and the ``staggered'' one $ D_l =  (-1)^l J \tan\delta$.
In both cases the Hamiltonian is reduced to the $XXZ$ model ~:

        \begin{eqnarray}
        {\cal H} &=& \sum_{l=1}^L(
         J^x
        ( {\widetilde S}^x_l {\widetilde S}^x_{l+1} +
        {\widetilde S}^y_l {\widetilde S}^y_{l+1})
        +  J {\widetilde S}_l^z {\widetilde S}_{l+1}^z )
        \label{newHam}
        \end{eqnarray}
with $J^x = \sqrt{J^2+D_l^2}$ independent of $l$.

Our subsequent discussion is based on the observation, that the DM
interaction results in two effects for the observable susceptibility of
the system.  First effect is the modification of the spectrum, as seen
in the equivalent Hamiltonian (\ref{newHam}).  The appearance of the
"easy-plane" anisotropy ($J^x>J$), however, does not lead to a gap in
the spectrum.  The exact solution of (\ref{newHam}) for $S=1/2$ shows
\cite{Lu-Pe} that the correlation functions $\langle
{\widetilde S}^x_l {\widetilde S}^x_m
\rangle \sim |l-m|^{-\nu}$ and $ \langle {\widetilde S}^z_l {\widetilde
S}^z_m \rangle \sim |l-m|^{-1/\nu}$ with $\nu = 1 - |\delta|/\pi$.
Since the value of DM exchange is expected to be small, $|D_l|\ll J$,
the long-distance decay of the above correlation functions is described
to a good accuracy by the ``isotropic'' Heisenberg situation, with $\nu
=1$. The second effect of DM interaction for the observables is the
explicit dependence of the relation (\ref{newspins}) between the new
and old spin variables on the values of $D_l$.  We focus our attention
below on the latter effect, which leads to the qualitative changes in
the experimentally observable susceptibilities.


The two-time Green's function for the operators $A$ and $B$  is
defined as
$        \chi_{AB}(t) =
        -i\theta(t)\langle [A(t), B] \rangle
$
where $[\ldots,\ldots]$ stands for a commutator and $\theta(t)=1$ at
$t>0$.

Upon the "twist" $e^{iU}$ the $z$-component of spin operators remains
unchanged, and one has for the longitudinal $zz$ susceptibility
$\chi^{zz}_{lm}(t) =
-i\theta(t)\langle [S^z_l(t), S^z_m] \rangle = -i\theta(t)\langle
[{\widetilde S}^z_l(t), {\widetilde S}^z_m] \rangle \equiv {\cal
G}^{\|}_{lm}(t) $.  Therefore the observable $\chi^{zz}_{lm}(t)$ has a
commensurate antiferromagnetic modulation.

The expressions for the transverse spin susceptibility
are more complicated. It is convenient to introduce the
matrix \cite{Zvyagin}

        \begin{equation}
        \chi^\perp_{lm}(t) = -i\theta(t)\left [
        \begin{array}{cc}
        \langle [S^x_l(t), S^x_m] \rangle , &
        \langle [S^x_l(t), S^y_m] \rangle \\
        \langle [S^y_l(t), S^x_m] \rangle , &
        \langle [S^y_l(t), S^y_m] \rangle
        \end{array}
        \right],
        \end{equation}

\noindent
in the initial system (\ref{iniHam}).
In the simpler "twisted" system (\ref{newHam}) we have
$
-i\theta(t)\langle [{\widetilde S}^x_l(t), {\widetilde S}^x_m] \rangle =
-i\theta(t)\langle [{\widetilde S}^y_l(t), {\widetilde S}^y_m] \rangle
\equiv {\cal G}^{\perp}_{lm}(t)
$
and
$
\langle [{\widetilde S}^x_l(t), {\widetilde S}^y_m] \rangle =
\langle [{\widetilde S}^y_l(t), {\widetilde S}^x_m] \rangle =0
$.
Returning back to quantities
$\langle [S^\alpha_l(t), S^\beta_m] \rangle$
with the use of (\ref{newspins}), we get

        \begin{equation}
        \chi^\perp_{lm}(t) =
        {\cal G}^{\perp}_{lm}(t)
        \left [
        \begin{array}{cc}
        \cos \alpha_{l,m},& -\sin \alpha_{l,m} \\
        \sin \alpha_{l,m},& \cos \alpha_{l,m}
        \end{array}
        \right]
        \label{corr-tw}
        \end{equation}
with $\alpha_{l,m} = \alpha_l - \alpha_m$.

For later comparison, it is worth to consider first the case of
the staggered DM interaction. \cite{Aff-Osh} We have
$D_l = (-1)^l J \tan \delta$ and $\alpha_{l,m} = ((-1)^l -
(-1)^m) \delta/2$. In this case we write $\cos \alpha_{l,m} =
\cos^2(\delta/2) + (-1)^{l-m} \sin^2(\delta/2) $ and $\sin \alpha_{l,m}
= ((-1)^l - (-1)^m) (\sin \delta)/2$. Clearly, the off-diagonal
components $\chi_{lm}^{xy}(t)$ of the matrix (\ref{corr-tw}) do not
depend on the difference $(l-m)$ only and the
two-momenta Fourier transform $A(q,q',\omega) = \int dt\sum_{lm}
e^{iql-iq'm-i\omega t} A_{lm}(t)$ should be introduced.
Then we obtain the off-diagonal
components in the form $\chi^{xy}(q,q',\omega) = ( {\cal
G}^\perp(q,\omega) - {\cal G}^\perp(q-\pi,\omega) ) \sum_\tau
\delta(q-q'-\pi+\tau)$ where $\sum_\tau$ stands for the sum over all
vectors $\tau = 2\pi n$ of the reciprocal lattice. However, apparently
in all physical observables one finds the symmetrized form of the
susceptibility ($q=q'$) and the off-diagonal terms in the matrix
$\chi^\perp$ vanish.  Therefore in the case of staggered DM interaction
one is left with the diagonal component of the matrix
$\chi^\perp(q,\omega)$ of the form ~:

        \begin{eqnarray}
        \chi^\perp(q,\omega) &=&
        {\cal G}^\perp(q,\omega)
        \cos^2(\delta/2)
        \label{corr-stag} \\
        && +
        \frac12 [{\cal G}^\perp(q-\pi,\omega) +
        {\cal G}^\perp(q+\pi,\omega)]
        \sin^2(\delta/2) \nonumber
        \end{eqnarray}

We see that the regions of the AF and the ferromagnetic
fluctuations are mixed in the observable susceptibility.
A consequence of this feature \cite{Aff-Osh} is the anomalous
temperature behavior of the uniform static susceptibility
$\chi^{xx}(0,0) = \chi^{yy}(0,0)$ for the AF chain (see also the
discussion after Eq.(5.3) in the original Moriya's paper
\cite{DzyaMo}).  It is known \cite{Schulz-temp} that in the
Heisenberg $S=1/2$ chain one has ${\cal G}^\perp(0,0) \sim
J^{-1}$ and ${\cal G}^\perp(\pi,0) \sim T^{-2+\nu}$, therefore
$\chi^{zz}(0,0) \sim J^{-1} $ and

        \begin{eqnarray}
        \chi^{xx}(0,0) &=& \chi^{yy}(0,0)
        \sim J^{-1}[ const + {\delta^2}(J/T)^{1+|\delta|/\pi}]
        \label{susc-stag}
        \end{eqnarray}

\noindent
The Eq.\ (\ref{susc-stag}) has a simple physical meaning.
Indeed, in the considered case the operator $U$ ``cants'' the local
coordinate frames by an angle $\pm \delta/2$. It leads effectively
to the non-compensated spin $\Delta S = \delta/4$ in the $x-y$ plane.
Being the spins $\Delta S$ free, it would then lead to the Curie law for
the susceptibility $\chi \sim \langle\Delta S^2\rangle/T$.
The 1D character of the interacting spin system results in the
nontrivial exponent in the $T-$dependence of this term (cf.\ also
\cite{Aff-Osh}).

At the same time, the above transfer of the spectral weight, Eq.\
\ref{corr-stag}, is apparently negligible to be observed in the
neutron scattering experiments.

On the other hand, for the ``uniform'' DM interaction $D_l = D$ we have
$\alpha_{l,m} = (l -m) \delta$. It results in the
incommensurability of the transverse spin correlations. Fourier
transforming Eq.\ (\ref{corr-tw}), we obtain

        \begin{eqnarray}
        \chi^\perp(q,\omega) &=&
        \frac12
        {\cal G}^\perp(q+\delta,\omega)
        \left [
        \begin{array}{cc}
        1 ,& i \\
        -i,& 1
        \end{array}
        \right]
        \nonumber \\ &&+
        \frac12
        {\cal G}^\perp(q-\delta,\omega)
        \left [
        \begin{array}{cc}
        1 ,& -i \\
        i,& 1
        \end{array}
        \right]
        \label{corr-uni}
        \end{eqnarray}


Let us discuss the physical consequences of this expression.
Evidently, $\chi^{xx}(0,0) \sim const$ in this case and the presence of
the ``uniform'' DM interaction is not revealed by the measurements of
the temperature dependence of the uniform static susceptibility.  Much
more interesting are the implications of (\ref{corr-uni}) for the
polarized neutron scattering experiments. The basic quantity here is
neutron scattering cross-section, which is connected to the Green's
function $\chi^{\alpha\beta}(q,\omega)$ of the spin system.  It is
convenient to write $\chi^{\alpha\beta} = \chi^{\alpha\beta}_S - i
\chi^{\alpha\beta}_A$, with the symmetric and antisymmetric tensors,
$\chi^{\alpha\beta}_S$ and $\chi^{\alpha\beta}_A$, respectively.  Up to
fundamental constants, we have \cite{Mal-chir} :

        \begin{eqnarray}
        \frac{d^2\sigma(q,\omega)}{d\Omega d\omega} &\sim&
        N(-\omega)
        [
        Im \chi^{\alpha\beta}_S(q,\omega)
        (\delta^{\alpha\beta} - \widehat Q^{\alpha}
        \widehat Q^{\beta} )
        \nonumber \\ &&
        +
        Im \chi^{\alpha\beta}_A(q,\omega)
        \epsilon_{\alpha\beta\gamma}
        \widehat Q^{\gamma}(\widehat {\bf Q} {\bf P}_0)
        ]
        ,
        \label{crosssection}
        \end{eqnarray}

\noindent
where $N(\omega)$ is the Planck function,
the unit vector $\widehat {\bf Q} = {\bf Q}/Q$ is directed along the
neutron's momentum transfer ${\bf Q}$,
$\epsilon_{\alpha\beta\gamma}$
is totally antisymmetric tensor, ${\bf P}_0$ is the incident
neutron's polarization and $q$ is the on-chain projection of
${\bf Q}$.

From (\ref{crosssection}) we see that if the whole crystal is
characterized by Dzyaloshinskii vector $\bf D$ (uniform situation, Eq.
\ref{corr-uni}), then the polarization-dependent part of cross-section
is non-zero and is given by

        \begin{eqnarray}
        \frac{d^2\sigma_1}{d\Omega d\omega} &\sim&
        ({\bf D}\widehat {\bf Q})(\widehat {\bf Q} {\bf P}_0)
        Im \frac{{\cal G}^\perp(q+\delta,\omega) -
        {\cal G}^\perp(q-\delta,\omega)}{2D}.
        \label{polariz}
        \end{eqnarray}
A certain subtlety should be discussed here. Under the parity
transformation we have $q\to-q$, ${\bf D}\to {\bf
D}$, ${\bf P}_0\to {\bf P}_0$. At the first glance $\delta
\to\delta$ and thus ${d^2\sigma_1}/{d\Omega d\omega}$ changes
the sign, as it should not be. An inspection of (\ref{canon}),
(\ref{corr-tw}) (\ref{corr-uni}) shows however that the quantity
$\delta$ in (\ref{polariz}) appears as the differential of
$\alpha_l$. The latter object is the sum of the phases
$\delta_j$ over the bonds $j$ to the left of $l$.
Hence we have under the parity transformation $\alpha_l \to
-\alpha_l + const$ and $\delta \to -\delta$ in (\ref{polariz}),
which restores the desired property of the cross-section.

The contribution of the symmetric part of
$\chi^{\alpha\beta}$ to ${d^2\sigma(q,\omega)}/{d\Omega d\omega}$
in the considered case of uniform ${\bf D}$ is two-fold. One
still has the commensurate fluctuations of spin components $S^z
\| {\bf D}$, with a peak at the AF position.  At the same time,
the DM interaction splits the AF peak related to transverse
fluctuations into two peaks of the weight $1/2$, Eq.\
\ref{corr-uni}. The relative weights of these two structures
depend on the direction of ${\bf Q}$ as follows

        \begin{eqnarray}
        \frac{d^2\sigma_2(q,\omega)}{d\Omega d\omega} &\sim&
        (1+ \widehat Q_z^2)\,
        Im \frac{{\cal G}^\perp(q+\delta,\omega) +
        {\cal G}^\perp(q-\delta,\omega)}{2}
        \nonumber \\ &&
        + (1- \widehat Q_z^2)\,
        Im\, {\cal G}^\|(q,\omega)  .
        \label{unpolariz}
        \end{eqnarray}

\noindent
Note that at low temperatures,
the incommensurate peaks have more singular behavior according to
our discussion after Eq.\ (\ref{newHam}).

An important thing to be stressed here is the following. It is known
that the incommensurate long-range magnetic structures may arise due to
the competing interactions in the spin system. In this case one expects
that all three diagonal components of the spin susceptibility
$\chi^{\alpha\alpha}$ are peaked {\em in the paramagnetic region} at
the same incommensurate wave-vector. The off-diagonal components of
$\chi^{\alpha\beta}$ are absent. This is fairly different from the
picture described above, Eqs.\  \ref{polariz}, \ref{unpolariz}.
Hence the experimental observation of the incommensurability phenomenon
in the paramagnetic phase, accompanied by the polarization dependence
of the neutron scattering cross-section could serve as an indication
to the presence of the uniform DM interaction. Remarkably, the value
and the direction of the pseudo-vector ${\bf D}$ can be, in
principle, determined this way.

In reality, however, the macroscopic sample is rarely uniform and it
should be expected to split to domains with different directions of
${\bf D}$.  To account for this situation, it is instructive to analyze
a model where the value of the DM interaction $D_l$ takes randomly two
values $\pm J \tan\delta$.

Consider first the oversimplified case when $\dla D_l\dra = 0$ and
$\dla D_l D_m \dra= 0$ for $l\neq m$, here $\dla\ldots\dra$ denotes
averaging over the realizations. In
this case the spectrum is still defined by Eq.(\ref{newHam}), and
$\chi^{zz}$ is given by the above expression. At the same time, one
can easily show that the averaged susceptibility $\chi^\perp$ has a
diagonal form and exhibits an exponential decay of correlations :

        \begin{equation}
        \dla\chi^\perp_{lm}(t)\dra =
        {\cal G}^{\perp}_{lm}(t)
        \exp(-|l-m|/l_\ast)
        \label{corr-ran}
        \end{equation}
with the correlation length $l_\ast = -1/\ln(\cos\delta) \sim
\delta^{-2}$.

Now consider a more realistic situation when one still has $\delta_j =
\pm \delta$, but the signs of $\delta_j$ on the adjacent bonds are
correlated. The diagonal and off-dia\-go\-nal parts of the matrix
(\ref{corr-tw}) are given, respectively, by the real and imaginary part
of the average

        \begin{equation}
        \dla \exp i\alpha_{l,m}\dra \equiv
        \sum_{\{\delta_j \}}
        p(\delta_1,\ldots,\delta_L)
        \exp(i\sum_{k=m}^{l-1} \delta_k).
        \label{average}
        \end{equation}

\noindent
We assume that the joint distribution function
$p(\delta_1,\ldots,\delta_n)$
has a Markovian character,
$p(\delta_1,\ldots,\delta_n) =
p(\delta_1, \ldots, \delta_{n-1}) \hat p (\delta_n|\delta_{n-1})$.
In this physically important case we arrive at the dichotomous Markovian
noise $\delta_j$ with a discrete ``time'' $j$. \cite{noise}
We set $\dla\delta_j\dra = \delta d$ which defines
the on-site (``equilibrium'')
probability as $p_0 = (\frac{1+d}2,\frac{1-d}2)$.
The matrix $\hat p (\delta_n|\delta_{n-1})$ satisfies the
``conservation laws'' for the total and equilibrium probabilities,
$(1,1)\cdot \hat p (\delta_n|\delta_{n-1})=(1,1)$ and
$\hat p (\delta_n|\delta_{n-1}) \cdot p_0=p_0$, respectively. These
equalities fix $\hat p (\delta_n|\delta_{n-1})$ in the form

        \[
        \hat p =
        \left[\begin{array}{cc}
        1-x(1-d),& x(1+d)  \\
        x(1-d),& 1-x(1+d)
        \end{array}\right]
        \]
for all $n$, which corresponds to the following
correlator on the adjacent sites ~:
$\dla\delta_j
\delta_{j+1}\dra - \dla~\delta_j~\dra^2= \delta^2 (1-d^2)(1-2x)$.
The latter equalities mean
that the absence of correlations corresponds to $x=1/2$
and the correlation lengths for positive and negative sequences of
$\delta_j$ are $1/l_{1,2} = x(1 \mp d)$. Introducing the matrix
${\cal D}=diag(e^{i\delta},e^{-i\delta})$, the
quantity (\ref{average}) is represented as a product $(1,1)\cdot
({\cal D}\cdot \hat p)^{l-m-1}\cdot{\cal D}\cdot  p_0$, which
is evaluated using the multiplication rules for the Pauli matrices.
After straightforward, though tedious, calculation, we obtain the
average (\ref{average}) in general form, which is somewhat simplified
in two principal cases of small $\delta$ ~: i) $\delta \ll x\sim1$ and
ii) $\delta\sim x \ll1$.  In the first case, keeping the terms of order
of $\delta^2$, we have

        \begin{eqnarray}
        \dla e^{ i\alpha_{m+n,m}}\dra
        &\simeq&
        \exp[n(i\delta d -\delta^2 a_1) +\delta^2 a_2]
        \label{avg-xsim1}
        \\ &-&
        \delta^2 a_2
        (1-2x)^{n}
        \exp[(n-1)(-i\delta d +\delta^2 a_1)]
        \nonumber
        \end{eqnarray}
here $a_1 = (1-x)(1-d^2)/(2x)$ and $a_2 = (1-d^2)(1-2x)/(4x^2)$.
We see that the incommensurability wave vector is defined by the
average on-bond value $\delta d$.
Note that
Eqs.\ (\ref{corr-stag}), (\ref{corr-ran}) are recovered
at $d=0$, $x=1$ and $d=0$, $x=1/2$, respectively.

When $x\sim \delta \ll1$, we come to a more complicated situation.
We have in this case

        \begin{eqnarray}
        \dla e^{ i\alpha_{m+n,m}}\dra
        &\simeq&
        e^{-x n}
        \left[\cosh bn + \frac{\sinh bn}{b}(x+ i\delta d)
        \right]
        \label{avg-x<<1}
        \end{eqnarray}
with $b = \sqrt{x^2+2ix\delta d- \delta^2}$.
We return to Eq.\ (\ref{corr-uni}) at $x=0$ and $d=1$.

It should be stressed that at $d=0$ one has $Im \dla e^{
i\alpha_{m+n,m}}\dra =0$ both in Eqs.\ (\ref{avg-xsim1}),
(\ref{avg-x<<1}) and in general case. It corresponds to the fact that
the off-diagonal components of susceptibility $\chi^\perp$, Eq.\
(\ref{corr-tw}), vanish in the system with zero average Dzyaloshinskii
vector $\dla {\bf D} \dra$.
As a result,
the observable cross-section is polarization-independent.
The position of maximum of the transverse spin fluctuations, though,
may be incommensurate one for the almost uniform DM interaction,
as seen from (\ref{avg-x<<1}) at $d=0$ and $x\to 0$.

Now we discuss the possible generalization of our approach to a higher
dimensional case.  Consider a planar system with spins ${\bf S}_{lm}$
labeled by two indices.  The interaction between spins takes place in two
directions, and we write the corresponding quantities as $J^{(\alpha)}_{lm}$
and ${\bf D}^{(\alpha)}_{lm}$, with $\alpha=x,y$.  For simplicity we
consider the case when the vectors ${\bf D}^{(\alpha)}_{lm}$ lie along one
direction, with possible variation in their magnitude.  We introduce then
two angles, $\delta^x_{lm}= \tan^{-1} (D^{(x)}_{lm}/J^{(x)}_{lm})$ and
$\delta^y_{lm}= \tan^{-1} (D^{(y)}_{lm}/J^{(y)}_{lm})$.
We are interested to arrive to a symmetrized
Hamiltonian, similar to (\ref{newHam}), by making the
transformation with $U = \sum_{l,m=1}^L \alpha_{lm} S_{lm}^z$.
One can show that this transformation is possible if and only if
$\delta^x_{l,m+1} - \delta^x_{lm} = \delta^y_{l+1,m} -
\delta^y_{lm}$. In this case $\alpha_{lm}$ is uniquely defined
and may be written in a form $\alpha_{lm} =
\sum_{j=1}^{l-1}\delta^x_{j,1} +\sum_{j=1}^{m-1}\delta^y_{l-1,j}
$.  The latter relations are not surprizing when we note that in
the continuum limit they read as $\nabla \times {\vec
\delta}_{\bf r} = 0$, while ${\vec \delta}_{\bf r} =
(\delta^x_{\bf r}, \delta^y_{\bf r}) = \nabla
\alpha_{\bf r}$.  In particular, the two-dimensional situation
with $D^{(x)}_{lm} = D$ and $D^{(y)}_{lm} = 0$ allows this
transformation.
The calculation of observables and the generalization
to a three-dimensional case are done in the way similar to the
above one.

In conclusion, we discuss the observables in the spin system with the
antisymmetric DM interaction. We show that
in one spatial dimension the exactly found spectrum of such problem
may coincide for different choices of the parameters of DM interaction.
Despite this fact, the observable spin-spin correlation functions may
differ crucially, and we discuss this feature with the application to
the neutron scattering experiments. In particular,
the incommensurability and the polarization
dependence of the neutron scattering may be used for the determination
of the value of the uniform DM interaction.

We thank S.L.\ Ginzburg, D.R.\ Grempel, A. Gukasov,
D. Petitgrand, V.P.\ Plakhty for useful discussions and comments.  This
work was supported by Russian State Program for Statistical Physics
(Grant VIII-2), RFBR Grant No.  00-02-16873, and the Russian Program
"Neutron Studies of Condensed Matter".


\end{multicols}
\end{document}